\documentclass[twocolumn,final,showpacs,
               showkeys,
               prl,superbib,
               letterpaper,balancelastpage,
               lengthcheck,byrevtex,bibnotes,
               footinbib,
               ]{revtex4}
\usepackage{graphics}
\usepackage{dcolumn}
\usepackage{graphicx,natbib,amssymb}
\usepackage[thinspace,squaren]{SIunits}

\def\etc{{\em etc.}}

\def\full{\protect\mbox{------}}
\def\kesik{\protect\mbox{--\, --\, --}}
\def\chain{\protect\mbox{-- $\cdot$ --}}
\def\dd{{\rm d}}

\begin{document}

\title{Extracting spectral density function of a binary composite without {\em a-priori} assumption}
\author{\firstname Enis \surname Tuncer}
\email{enis.tuncer@physics.org}
\affiliation{Applied Condensed-Matter Physics, Department of Physics, University of Potsdam, D-14469 Potsdam Germany}
\date{\today}

\begin{abstract}
The spectral representation separates the contributions of geometrical arrangement (topology) and intrinsic constituent properties in a composite. The aim of paper is to present a numerical algorithm based on the Monte Carlo integration and contrainted-least-squares methods to resolve the spectral density function for a given system. The numerical method is verified by comparing the results with those of Maxwell-Garnett effective permittivity expression. Later, it is applied to a well-studied rock-and-brine system to instruct its utility. The presented method yields significant microstructural information in improving our understanding how microstructure influences the macroscopic behaviour of composites without any intricate mathematics.
\end{abstract}
\keywords{Spectral representation, composites, dielectric permittivity, micro-structural information}
\pacs{77.22.-d, 77.22.Ch, 02.70.Hm, 02.70.Uu, 05.10.Ln, 07.05.Kf, 61.18-j}
\maketitle


Theory of mixtures and their electrical properties have attracted researchers to seek a relation between intrinsic properties of the parts forming the mixture  (constituents) and their spatial arrangement inside the mixture\cite{Landauer1978}. Bergman\cite{Bergman1978} has proposed a mathematical way for representing the effective dielectric permittivity $\varepsilon_{\sf e}$ of a binary mixture as a function of permittivities of its constituents, $\varepsilon_1$ and $\varepsilon_2$, and an integral equation, which includes the geometrical contributions. It is called {\em the spectral density representation} (SDR). After the introduction of non-destructive measurement techniques and systems, such as electrical\cite{Jonscher1983,McCrum} or acoustic impedance spectroscopy\cite{McCrum}, the impedance of materials (either pure or composite) could be recorded for various frequencies $\nu$. Then, the frequency could be used as a probe to obtain microstructural information with the application of the SDR\cite{Stroud,GhoshFuchs,Stroud1999,Day,Lei2001,Gonc2003,Cherkaev2003}. This can only be achieved if ({\em i}) no influence of $\nu$ on the geometrical arrangement of phases is present\cite{footnote1}, and ({\em ii}) the intrinsic properties of phases are known as a function of $\nu$. Numerical\cite{Day,Gonc2003,Cherkaev2003} and analytical\cite{Stroud,GhoshFuchs,Stroud1999} approaches have been used and proposed to resolve the spectral density function (SDF) for composites. Although numerical approaches could be prefered over the analytical ones, which are emprirical expressions and are not universal, they solve a nontrivial---ill-posed---inverse problem\cite{Cherkaev2003}. Here, we apply a recently developed numerical method\cite{Tuncer} to extract the SDF of a binary mixture. The method is based on the Monte Carlo integration and constrained-least-squares (C-LSQ) algorithms. By using this procedure the integration constant becomes continues rather than discrete as in regulation algortihms. First, the verification of the proposed method is presented by considering the Maxwell-Garnett (MG) effective dielectric function\cite{Maxwell_Garnett}. Later, it is applied to the dielectric data of a rock-and-brine system\cite{Kenyon}, which has been also used by Refs.\onlinecite{Stroud} and \onlinecite{GhoshFuchs} to test their analytical expressions.

\begin{figure}[t]
  \centering
  \includegraphics[]{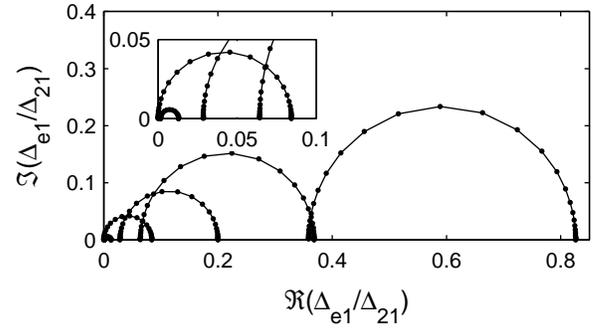}
  \caption{Parametric plot of the scaled mixture permittivity. The symbols are the analytical model of Maxwell-Garnett equation, and the solid lines (\full) are the values calculated from the spectral functions obtained from the proposed numerical method.  The semi-circles from large to small corresponds to $q_2=\{0.95,\, 0.7,\,0.5,\, 0.3,\,0.05\}$, respectively. The inset is the enlargement of the values close to the origin for $q_2=\{0.05,\,0.30\}$.}
  \label{fig:normalized}
\end{figure}

For a binary composite system with constituent permittivities $\varepsilon_1$ and $\varepsilon_2$, and concentrations $q_1$ and $q_2$, ($q_1+q_2=1$), and with an effective permittivity $\varepsilon_{\sf e}$, the SDR is expressed as\cite{footnote2},
\begin{eqnarray}
  \label{eq:1}
  \Delta_{{\sf e}i}/\Delta_{ji} - A_j = \int_0^1 {\sf g}_j(x)\,[1+\varepsilon_i^{-1}\Delta_{ji}x]^{-1}\, \dd x
\end{eqnarray}
where, $\Delta_{ij}=\varepsilon_i-\varepsilon_j$, and is complex and frequency dependent. $A_j$ is a constant, and depends on the concentration and structure of the composite. The SDF is ${\sf g}(x)$, and it is sought by the presented procedure. The SDF satisfies $\int{\sf g}_j(x)\dd x=q_j$\cite{Bergman1978,Gonc2003} and $\int x {\sf g}_j(x)\dd x=q_j \, q_i/d$, where $d$ is the dimension of the system. The shape of the inclusions in a matrix can also be related to $d$\cite{Sillars1937}. Finally,  $x$ is called the depolarization factor.

The numerical procedure is briefly as follows: first the integral in Eq.~(\ref{eq:1}) is written in a summation form over some number of randomly selected (known) $x_n$-values, $x_n\in [0,1]$. This  converts the non-linear problem in hand to a linear one with ${\sf g}_{j_{n}}$ being unknowns. Later, a C-LSQ is applied to get the corresponding ${\sf g}_{j_{n}}$-values:
\begin{eqnarray}
  \label{eq:2}
  \min||{\bf \Delta}-{\bf K}{\sf g}_{j_{n}} ||_2 \quad {\rm and} \quad {\sf g}_{j_{n}}\ge 0
\end{eqnarray}
where ${\bf \Delta}$ is the left-hand-side of Eq.~(\ref{eq:1}), and ${\bf K}$ is the kernel-matrix, $[1+\varepsilon_i^{-1}\Delta_{ji} x_n]^{-1}$. When this minimization is run over-and-over with new sets of $x_n$-values, most probable ${\sf g}_{j_{n}}$-values are obtained. For a large number of minimization loop, actually the $x$-axis becomes continues---the Monte Carlo integration hypothesis. Finally, the weighted distribution of ${\sf g}_{j_{n}}$ versus $x_n$ leads ${\sf g}(x)$\cite{footnote3}.  

Application of the numerical procedure to the MG expression should yield delta function distributions for ${\sf g}(x)$\cite{Stroud,Gonc2003}. The dielectric function for a $d$-dimensional (or composite with arbitrary shaped inclusions) MG composite is defined as 
\begin{eqnarray}
  \label{eq:4}
  \varepsilon_{\sf e}=\varepsilon_1 [1+ d\, q_2\, \Delta_{21}\,
  (q_1\,\Delta_{21} + d\, \varepsilon_1)^{-1}].
\end{eqnarray}
The resulting SDF is then,
\begin{eqnarray}
  \label{eq:3}
  {\sf g}_j (x)=\delta[x-(1-q_j)/d].
\end{eqnarray}
We choose the following values for dielectric functions of the phases: $\varepsilon_1=1-\imath\,(100\varepsilon_0\omega)^{-1}$ and $\varepsilon_2=10-\imath\,(\varepsilon_0\omega)^{-1}$ with $\omega=2\pi\nu$ and $\varepsilon_0=8.854\ \pico\farad\per\meter$. The left-hand-side of Eq.~(\ref{eq:1}) without the constant $A_2$ is plotted for a 3-dimensional composite ($d=3$ which corresponds to spherical inclusions) in Fig.~\ref{fig:normalized} as a parameteric plot of the imaginary part of $\Delta_{{\sf e}1}/\Delta_{21}$ against its real part. The graph is a semi-circle for the MG expression. In the figure, five different concentration levels are plotted, $q_2=\{0.05,\,0.3,\,0.5,\,0.7,\,0.95\}$, the inset shows the enlargement close to the origin, which illustrates the low concentrations, $q_2=\{.05,\,0.3\}$. The size of the semi-circles are proportional to the concentration of Phase 2. The analyses performed on the scaled effective permittivity, Eq.~(\ref{eq:1}), with the help of the applied method yield the solid lines (\full) in the figure.

\begin{figure}[t]
  \centering
  \includegraphics[]{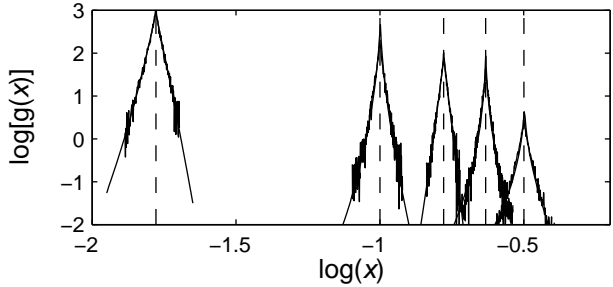}
  \caption{Calculated spectral density distributions, which correspond to delta sequences. The spectral functions from left to right corresponds to $q_2=\{0.95,\, 0.7,\, 0.5,\, 0.3,\,0.05\}$, respectively. The correcponding (calculated) $A_2$ values are $\{0.002,\,0.012,\,0.029,\,0.064,\,0.358\}$, respectively, for the considered concentrations. The dashed lines (\kesik) show the positions of the actual delta-functions for the MG expression.}
  \label{fig:gxMG}
\end{figure}

The corresponding ${\sf g}(x)$ are plotted in Fig.~\ref{fig:gxMG} on a log-log scale. In the figure, the expected locations of ${\sf g}(x)$ from Eq.~(\ref{eq:3}) are also shown with dashed lines (\kesik). The ${\sf g}(x)$-distributions obtained are analized by the L{\'e}vy distribution\cite{Levy}, which generates a delta-squence\cite{Butkov}. The solid lines (\full) illustrate the appropriate L{\'e}vy distributions. Various parameters from the statistical analyses and their expected values are presented in Table~\ref{tab:1}. The concentration values, $\overline{q_2}$, calculated from the integration of ${\sf g}(x)$ without {\em a-priori} assumption are $<1\%$ for the considered higher concentrations, and it is around 5\% for the lowest concentration, $q_2=0.05$. The localization parameter for the depolarization factor $x$, which is the most propable depolarization value, can be calculated by the integration of $[1-{\sf g}(x)]/d$ or with the help of statistical analysis. The estimated depolarization factors $\overline{x}$ are within $<1\%$ of the actual values stated by the proposed analytical expression\cite{Stroud,Gonc2003}. Finally, the product of the concentrations $\overline{q_1 q_2}$, the integration of $3x{\sf g}(x)$, calculated have also very good aggreement with those values expected from the definitions of the SDR.

\begin{table}[pt]
\squeezetable
  \caption{Comparison between the reults of the proposed numerical approach and those of the  L{\'e}vy statistics and the given analytical SDF for the MG effective permittivity expressions for various concentrations. The bars on the quantities indicate that they are calculated from the numerical results.}
  \centering
\begin{ruledtabular}
    \begin{tabular}{cccccccc}
$q_2$   & 
$\overline{q_2}$ \footnotemark[1] &  
$\overline{x}$ \footnotemark[2] & 
$q_1/d$ \footnotemark[3] & 
$A_2{_{in}}$ \footnotemark[4] &
$A_2{_{out}}$ \footnotemark[5] &
$\overline{q_1 q_2}$ \footnotemark[6] &
$q_1 q_2$ \footnotemark[3]\\ 
      \colrule
      0.05  & 0.053 & 0.318 & 0.316 & 0.002 & 0.002 & 0.057 & 0.048\\
      0.30  & 0.301 & 0.234 & 0.233 & 0.012 & 0.013 & 0.213 & 0.210\\
      0.50  & 0.050 & 0.167 & 0.167 & 0.029 & 0.029 & 0.249 & 0.249\\
      0.70  & 0.704 & 0.100 & 0.100 & 0.064 & 0.064 & 0.213 & 0.280\\
      0.95  & 0.951 & 0.017 & 0.017 & 0.358 & 0.359 & 0.051 & 0.048\\
    \end{tabular}
\end{ruledtabular}
\footnotetext[1]{Calculated using the resulting ${\sf g}_2(x)$. Known from the definition of ${\sf g}_j(x)$---integral $\int_0^1{\sf g}_j(x)\dd x$ is equal to this value.}
  \footnotetext[2]{The localization parameter for the calculated L{\'e}vy distribution. The shape parameters and the amplitude of the L{\'e}vy distributions are disregarded.}  
\footnotetext[3]{Known from the definition of the SDF for the MG expression, Eq.~(\ref{eq:3}).}
\footnotetext[3]{$A_2$-value calculated before the numerical procedure using Eq.~(\ref{eq:1}).}
\footnotetext[4]{Mean $A_2$-value calculated during each Monte Carlo integration step in  the numerical procedure, Eq.~(\ref{eq:2}).}
\footnotetext[5]{Calculated using the resulting ${\sf g}_2(x)$ and $x$-values. Known from the definition of ${\sf g}_j(x)$---the values is equal to the integral $\int_0^1 3x{\sf g}_j(x)\dd x$.}
\label{tab:1}
\end{table}

\begin{figure}[t]
  \centering
  \includegraphics[]{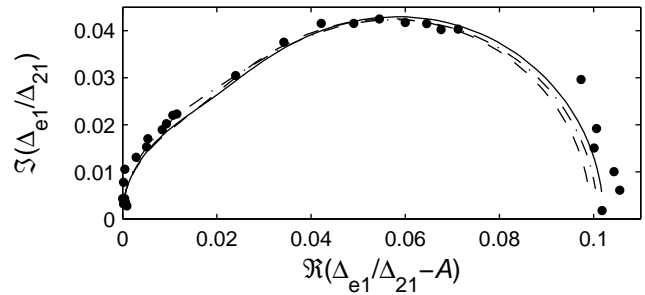}
  \caption{Parametric plot of the scaled rock-and-brine permittivity. The symbols ($\bullet$) are the experimental data of Ref.\onlinecite{Kenyon}. The chain line (\chain) is the results for the same assumptions as Refs.\onlinecite{Stroud} and \onlinecite{GhoshFuchs}. The solid (\full) and dashed (\kesik) lines are results obtained by two different water and composite conductivities; $\sigma_2=0.85\ \siemens\per\meter$ and $\sigma_{\sf e}=0.041\ \siemens\per\meter$ (\full), and $\sigma_2=0.85\ \siemens\per\meter$ and $\sigma_{\sf e}=0.038\ \siemens\per\meter$ (\kesik). 
}
  \label{fig:GFde}
\end{figure}

\begin{figure}[t]
  \centering
  \includegraphics[]{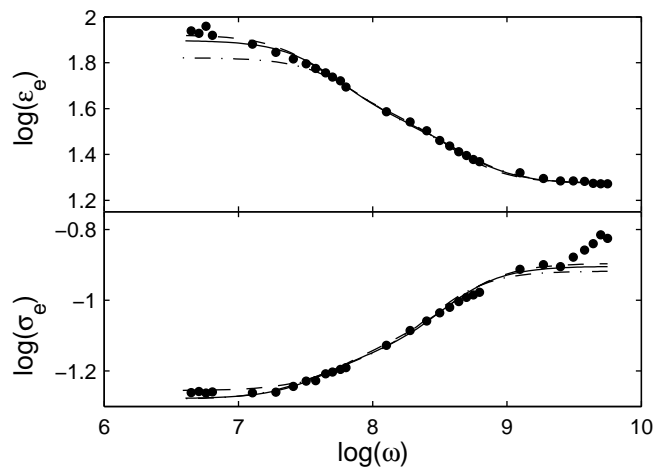}
  \caption{Measured ($\bullet$ from Ref.~\onlinecite{Kenyon}) and re-calculated dielectric permittivity $\Re(\varepsilon_{\sf e})$ and alternating current conductivity $\sigma_{\sf e}=\Im(\varepsilon_{\sf e}\varepsilon_0\omega)$. The chain line (\chain) is the results for the same assumptions as Refs.\onlinecite{Stroud} and \onlinecite{GhoshFuchs}. The line legends are the same as in Fig.~\ref{fig:GFde}. 
}
  \label{fig:GFes}
\end{figure}

We also test our procedure on a rock-and-brine composite system\cite{Kenyon}, which has been studied by various scientists\cite{Stroud,GhoshFuchs}. The same assumptions as in Ref.\onlinecite{Stroud} and \onlinecite{GhoshFuchs} are made to calculate the dielectric function of the brine (water-salt solution). The ohmic conductivity of the water is taken to be $\sigma_2=0.93\ \siemens\per\meter$, later the dielectric function of the brine at $T=75\ \degreecelsius$ is calculated by the following expression\cite{Stroud,GhoshFuchs},
\begin{eqnarray}
  \label{eq:5}
\varepsilon_2'(T)&=&94.88-0.2317\,T+0.000217\,T^2\nonumber \\
S(T)&=&5.363\,[(T+7)(82\sigma_2)^{-1}-0.0123]^{-1.047}\nonumber \\
\varepsilon'_2(T,S)&=&[\varepsilon_w{_0}^{-1}+0.0417\,S(1000-S)^{-1}]^{-1}\nonumber\\
\varepsilon_2(T,S)&=&\varepsilon'_2(T,S)-\imath\sigma_2(\varepsilon_0\omega)^{-1}
\end{eqnarray}
The relative permittivity of the rock is taken to be constant without any imaginary part, $\varepsilon_1=7.5$. The resulting scaled dielectric quantity in Eq.~(\ref{eq:1}) is presented in Fig.~\ref{fig:GFde}. Similar to Fig.~\ref{fig:normalized}, a semi-circle-like shape is observed. The first analysis with the above considerations results in an unsatisfactory calculated $\varepsilon_{\sf e}$ as presented with the chain line (\chain) in Fig.~\ref{fig:GFes}.  The low frequency side ($\omega<30\ \mega\hertz$)  of the real permittivity has discrepancies. Therefore, the experience of the author regarding dielectric data analyses suggests that the measured values at the low frequencies do not particularly satisfy the Kramers-Kronig relations\cite{LL,Tuncer}. Consequently, the application of the Kramers-Kronig relations yield lower effective composite conductivity then the original data, $\sigma_{\sf e}=0.055\ \siemens\per\meter$. Therefore, two different conductivities are adopted $\sigma_{\sf e}=0.041$ and $0.038\ \siemens\per\meter$, while we keep the conductivity of the water constant and lower than the previous consideration $\sigma_2=0.85\ \siemens\per\meter$. With these parameters as inputs, the resulting effective permittivity values have better agreement with those of measurements. And if compared to the results of \textcite{Stroud} and \textcite{GhoshFuchs}, our values have less residual than theirs.

\begin{figure}[t]
  \centering
  \includegraphics[]{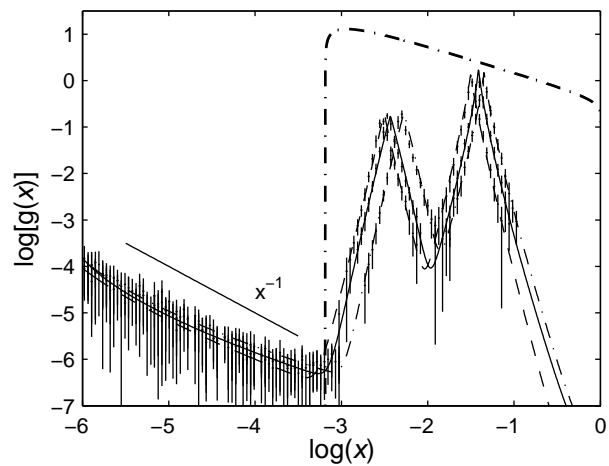}
  \caption{Calculated spectral density distributions. The lines represent the appropriate fitted L{\'e}vy distributions, and their legends are the same as in Fig.~\ref{fig:GFde}. The thick chain line is the SDF ${\sf g}(x)$ of Ref.\onlinecite{GhoshFuchs}. 
}
  \label{fig:gxGF}
\end{figure}

In Fig.~\ref{fig:gxGF}, the obtained ${\sf g}(x)$ are presented. It is striking that two very distinct peaks are observed whatever the initial assumptions for the conductivities of the water as well as the composite are. The ${\sf g}(x)$ can be divided into three sub-SDF, which are located around $x=\{0,\,0.004,\,0.04 \}$. It is clear that the original data can be modeled by only two SDF as delta sequences\cite{Lei2001,Gonc2003,Tuncer2001a} without a sophisticated mathematics. The SDF of Ref.\onlinecite{GhoshFuchs} is also displayed as a comparison with the thick chain line (\chain), which has been valuable to give limits for the depolarization factor $x$. However, in the case of Kenyon's data\cite{Kenyon} it overestimates the upper limit, which has been 1. The two peculiar depolarization factors resolved from the peaks have concentrations of $0.111$ and $0.023$, respectively, which are calculated from the L{\'e}vy distributions. The low $x$-side of ${\sf g}(x)$ yields a very small concentration ($\sim 10^{-5}$) for that particular depolarization process. If we take into account the yielding concentrations of the brine in the system ($q_2\approx 0.134$), we can state that the three peaks correspond to oblate to needle like porous structures of the brine with shape factor estimates $d\approx q_1/\overline{x}\approx\{10^5,\,200,\,1.2\}$ for $x=\{0,\,0.004,\,0.04\}$, respectively\cite{Sillars1937}. 
It is clear that the brine-phase forms channel-like structures, because of high $d$ values. Continuous percolating paths are formed when ${\sf g}(x)=\delta(x)$ and $d\rightarrow\infty$, which corresponds to a structure with channels parallel to field direction.

As a concluding remark, an effective numerical method is presented to extract the SDF of a binary composite system. It is tested on both `ideal' and measured dielectric data for composites.  The proposed method not only extracts the SDF, it also yields volume fractions of both constituents as well as the correponding depolarization processes and phase shapes even if they are not known at the beginning. It is shown that it can resolve unique individual depolarization processes, which could indeed be used to obtain valuable microstructural information regarding the composite and its constituents in various research fields, in which impedance spectroscopy is used for characterization of materials, such as, polymeric, pharmaceutical, biological, building, colloidal, porous, \etc

\printtables
\printfigures

\begin{thebibliography}{99}
\bibitem[{\citenamefont{}(2004)}]{footnote1}{The geometry should be static at each frequency, meaning that no piezoelectricity exists in the constituents, and the elastic properties of the phases should be the same or similar to each other otherwise there would be a nonzero displacement vectors (deformation) in the composite.}
\bibitem[{\citenamefont{}(2004)}]{footnote2}{Several different notations have been used in the literature\cite{Stroud,GhoshFuchs,Day}. Here, we rearrange the expression by Ref.\onlinecite{GhoshFuchs}, and obtain a similar one those used by Refs.\onlinecite{Stroud} and \onlinecite{Day}. 
}
\bibitem[{\citenamefont{}(2004)}]{footnote3}{This is achieved by dividing the $x$-axis in channels and averaging ${\sf g}_{j_{n}}$ in each channel.}
\end{thebibliography}
\end{document}